\documentclass[reprint,superscriptaddress,nofootinbib,amsmath,amssymb,aps,prl,onecolumn,notitlepage]{revtex4-1}

\usepackage[a4paper, margin=2.50cm]{geometry}
\usepackage{xcolor}
\usepackage{bbm}
\usepackage{graphicx}
\usepackage{fancyhdr, color}
\usepackage{nameref}
\usepackage{hyperref} 
\usepackage{amsmath,amsfonts,enumerate,amsthm,amssymb,bbm}
\usepackage{braket}
\usepackage{float}
\usepackage[caption = false]{subfig}
\usepackage{enumitem}
\usepackage{ulem}

\theoremstyle{plain}

\theoremstyle{definition}

\theoremstyle{remark}

\def\<{\langle}
\def\E{ {\cal E} }

\def\T{ {\cal T} }

\def\I{ \mathbb{1} }

\def\I{ \mathbbm{1} }

\def\>{\rangle}
\def\<{\langle}
\DeclareMathOperator{\Tr}{Tr}

\newcommand{\be}{\begin{equation}}
\newcommand{\ee}{\end{equation}}

\newcommand{\red}[1]{{\color[rgb]{0,0,0}{#1}}}

\begin{document}

\fancyhead[C]{\sc \color[rgb]{0.4,0.2,0.9}{Quantum Thermodynamics book}}
\fancyhead[R]{}

\title{The Coherent Crooks Equality}

\author{Z. Holmes}
\email{z.holmes15@imperial.ac.uk} 
\affiliation{Department of Physics, Imperial College London, SW7 2BW, United Kingdom}

\begin{abstract}

This chapter reviews an information theoretic approach to deriving quantum fluctuation theorems that was developed in~\cite{aberg,alvaro}.
When a thermal system is driven from equilibrium, random quantities of work are required or produced: the Crooks equality is a classical fluctuation theorem that quantifies the probabilities of these work fluctuations. 
The framework summarised here generalises the Crooks equality to the quantum regime by 
modeling not only the driven system but also the control system and energy supply that enables the system to be driven. 
As is reasonably common within the information theoretic approach but high unusual for fluctuation theorems, this framework explicitly accounts for the energy conservation using only time independent Hamiltonians.  
We focus on explicating a key result of~\cite{aberg}: a Crooks-like equality for when the energy supply is allowed to exist in a superposition of energy eigenstates states.
\end{abstract}

\maketitle

\thispagestyle{fancy}

\section{I. \ Introduction} \label{sec:sectionlabel}

Fluctuation theorems are a pillar of contemporary thermodynamics: they are generalisations of the second law of thermodynamics that probe the irreversibility of non-equilibrium processes. Specifically, they consider systems driven from equilibrium and establish exact relations between the resultant thermal fluctuations. The emergent field of quantum fluctuation theorems aims to generalise classical fluctuation relations to the regime in which quantum phenomena such as coherence and entanglement become relevant. For a general introduction to both classical and quantum fluctuation relations see references~\cite{flucchap,fluctreview1,fluctreview2,QFT,litrev}.

Fluctuation theorems can broadly be classified as `detailed' or `integral' theorems, by the fluctuating quantity considered and by the nature of the non-equilibrium processes involved. `Detailed' fluctuation theorems consider pairs of non-equilibrium processes (a `forwards' one and its time-reversed equivalent) and quantify the relative probability of thermal fluctuations in the two processes. `Integral' theorems consider a single non-equilibrium process and quantify averages, or statistical moments, of such fluctuating quantities. The classical Crooks equality~\cite{Crooks} is a \textit{detailed} fluctuation theorem quantifying the \textit{fluctuating work} done on a system that is \textit{isothermally driven by a change in Hamiltonian}. The Jarzynki equality~\cite{Jarz} emerges as the integral variant of the Crooks equality. 

A plethora of quantum Crooks and Jarzynski equalities have been proposed over the last decade. The simplest approach defines the work done on a closed system as the change in energy found by performing projective measurements at the start and end of the non-equilibrium process~\cite{Tasaki,Kurchan,TasakiCrooks}. In this case, the classical Crooks equality holds unaltered. Generalisations of this simple quantum Crooks equality have largely focused on extensions to open quantum systems~\cite{openingup1,openingup2}, protocols represented by generic quantum channels~\cite{albash, Manzano, Rastegin} and alternative quantum work definitions~\cite{defofquantumwork1,defofquantumwork2,defofquantumwork3}. The latter includes definitions utilising quasi-probabilities~\cite{weakmeasurements,Allahverdyan}, the consistent histories framework~\cite{consistenthistories} and the quantum jump approach~\cite{chap14, quantumjump1,quantumjump2}. Such extensions lead to deviations from the classical equality. 

The information theoretic approach has proven an effective means of incorporating quantum mechanical phenomena into thermodynamics. The field rose to prominence with a series of papers probing the impact of entanglement on the second law of thermodynamics~\cite{entanglementsecondlaw}, Landauer erasure~\cite{entanglementnegentropy}, the thermodynamic arrow of time~\cite{entanglementarrowoftime} and thermalisation~\cite{entanglementthermalisation}. These were followed by results investigating work extraction~\cite{workextractionaberg,catalyticcoherence, catalyticworkextraction}, generalisations of the second laws~\cite{2ndlaws,constraintsbeyondfreeenergy}, and general criteria for state conversion~\cite{ThermodynamicTransformationsandworkextraction,completestateinterconversion} in quantum systems. Much of this research~\cite{ThermodynamicTransformationsandworkextraction,2ndlaws,workextractionaberg,completestateinterconversion, constraintsbeyondfreeenergy, catalyticcoherence, catalyticworkextraction} used insights from the thermal operations framework~\cite{chap25,resourcetheory,resourcetheory1,resourcetheory3}, a resource theory for quantum states out of thermal equilibrium.

The purpose of this chapter is to explicate an information theoretic approach that uses insights from the thermal operations framework to incorporate quantum coherence into fluctuation theorems. In particular, we seek to explain the coherent Crooks equality, Eq.~\eqref{Eq: Quantum Crooks} in Section II, which was derived by Johan \AA berg in~\cite{aberg}. This is a Crooks-like equality for a system with an energy supply that is in a superposition of energy eigenstates. While our focus is on this result, \'{A}lvaro Alhambra et al. independently used a similar methodology in~\cite{alvaro} to generalise the Jarzynski equality and to investigate the consequences of introducing fluctuating work into the thermal operations framework. 

\section{II. Sketch of framework and equality}

The key feature that distinguishes the framework of~\cite{aberg} and~\cite{alvaro} from other quantum fluctuation theorems is the manner in which energy conservation is explicitly accounted for. Firstly, the global Hamiltonian $H$ is assumed to be time independent. This is in marked contrast to the classical setup, as well as most previous quantum fluctuation theorems~\cite{Tasaki,Kurchan,TasakiCrooks,openingup1,openingup2, weakmeasurements,consistenthistories,quantumjump1,quantumjump2,Allahverdyan}, which utilise explicitly time dependent Hamiltonians.  Secondly, the non-equilibrium process is driven by a unitary $V$ that is `strictly' energy conserving~\cite{catalyticworkextraction} in the sense that it commutes with the global Hamiltonian, $[H, V] = 0$. These two restrictions are used within the resource theoretic approach to quantum thermodynamics~\cite{resourcetheory,resourcetheory1,resourcetheory3} where they have proven an effective means of carefully tracking the evolution of energy and coherence\footnote{We use the term `coherence' in
the sense of a `superposition of states belonging to different energy
eigenspaces'.} of a quantum system. 

\begin{figure}
\begin{center}
\subfloat[]{\includegraphics[width = 2.7in]{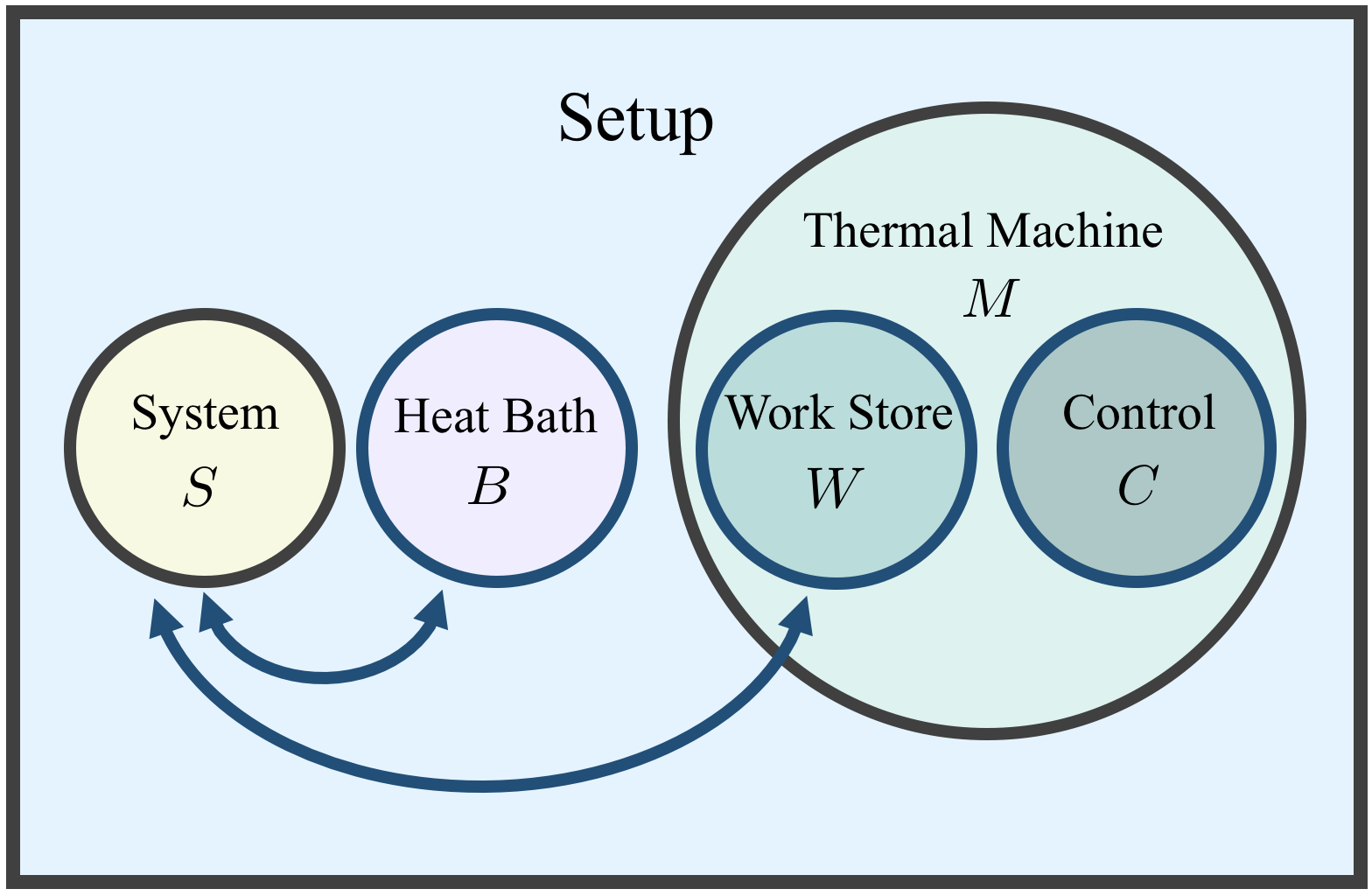}} \hspace{10mm}
\subfloat[]{\includegraphics[width = 2.7in]{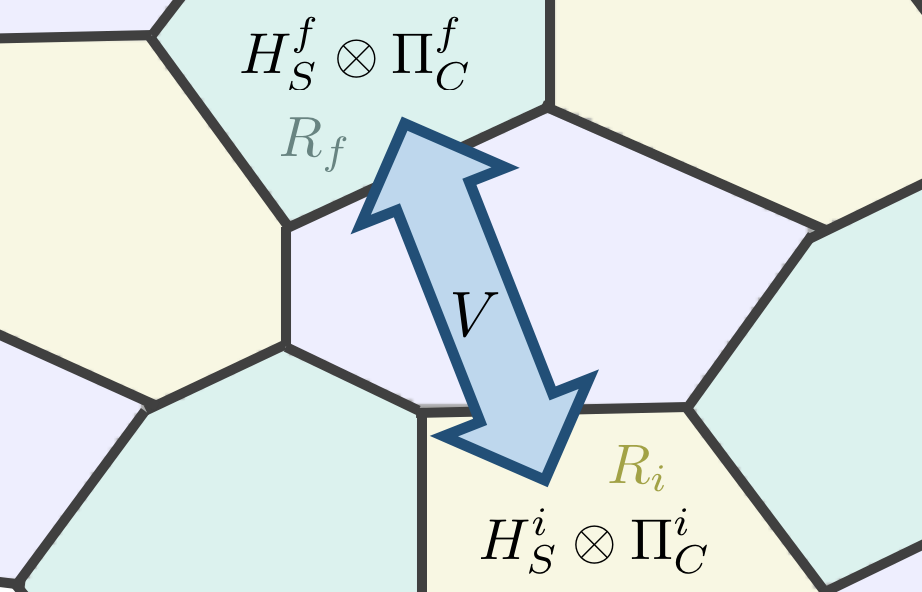}}
\caption{\label{Fig: sketch of equality plot}(a). A sketch of the setup consisting of a system ($S$), heat bath ($B$), work store ($W$), and control ($C$). This can be reduced to a bipartite system by using a single system that we call a 'thermal machine' ($M$) to act as both the work store and the control and by disregarding the heat bath and thinking of it as an implicit means of preparing the system in a thermal state. The arrows indicate the subsystems that exchange energy. (b). A sketch of a Hilbert space with a Hamiltonian that varies from region to region. An effectively time dependent Hamiltonian is induced if the support of the control evolves, under some unitary operation $V$, from one region to the next. In the forwards process the control evolves from region i to f. In the reverse, it evolves from region f to i.}
\end{center}
\end{figure}

In the standard fluctuation theorem setting a system is driven by a change in Hamiltonian resulting in exchanges of heat and work. This can be captured, using only time independent Hamiltonians and energy conserving processes, by explicitly modeling the subsystems that are implicitly involved. Specifically, the total setup is taken to consist of the:
\begin{itemize}
  \item System of interest $S$: this is the system that is driven by a change in Hamiltonian.
  \item Control system $C$: this enables the Hamiltonian of the system to be changed, $H_{S}^i \leftrightarrow H_{S}^f$.
  \item Work store $W$: this provides the energy required to change the Hamiltonian of the system. 
  \item Thermal bath $B$: this enables the system to relax in response to the change in its Hamiltonian. 
\end{itemize}

Crucially, the work store is assumed to be a quantum mechanical system and thus acts as a source of coherence. As the system in the standard Crooks equality starts in a thermal state which is diagonal in the energy eigenbasis, a means of introducing coherence, be it implicit or explicit, is required for any genuinely quantum mechanical Crooks equality.

\medskip

Having decided to explicitly model the work store, it is then additionally used in~\cite{aberg} to sidestep the challenges associated with how to define fluctuating quantum work (for a review see~\cite{defofquantumwork1, chap10}). The classical definition of fluctuating work~\cite{Crooks,Jarz,litrev} does not follow directly over to the quantum regime as quantum particles do not have well defined trajectories. Moreover, a recent no go theorem~\cite{defofquantumwork3} suggests that certain desirable features for a definition of quantum fluctuating work are incompatible. 

Rather than relying on a definition of `quantum work', the coherent Crooks equality is stated in terms of transition probabilities between work store states. This is a generalisation that is consistent with the classical approach. When the work store is prepared in an energy eigenstate, the transition probabilities agree with the standard two point measurement scheme~\cite{Tasaki,Kurchan,TasakiCrooks}. However, in the general case the work store can be prepared and found in a superposition of energy eigenstates.  

\medskip 
For the purpose of this chapter we will simplify the setup and present the coherent Crooks equality for a bipartite system. This is done by first treating the control system and work store as a single system, which we will call the thermal machine $M$, that has the dual role of driving the change in system Hamiltonian and providing or absorbing the energy required to do so. The equality is formulated in terms of  transition probabilities of the thermal machine. Secondly, we set the bath aside and think of it only as an implicit means of defining the temperature of the isothermal process and of preparing the system in a thermal state. The bipartite \red{setup} has the time-independent Hamiltonian,
\begin{equation}  \label{eq: total hamiltonian}
H_{MS} =H_M\otimes\I_S+\I_M\otimes H_S +H_{MS}^{\mbox{\tiny{int}}} \,
\end{equation}
comprised of the Hamiltonians $H_S$ and $H_M$ for the system and thermal machine, and their interaction $H_{MS}^{\mbox{\tiny{int}}}$. The full setup and this bipartite variant are sketched in Fig.~\ref{Fig: sketch of equality plot}a.

\begin{figure}
\begin{center}
\subfloat[Forwards Protocol]{\includegraphics[width = 2.7in]{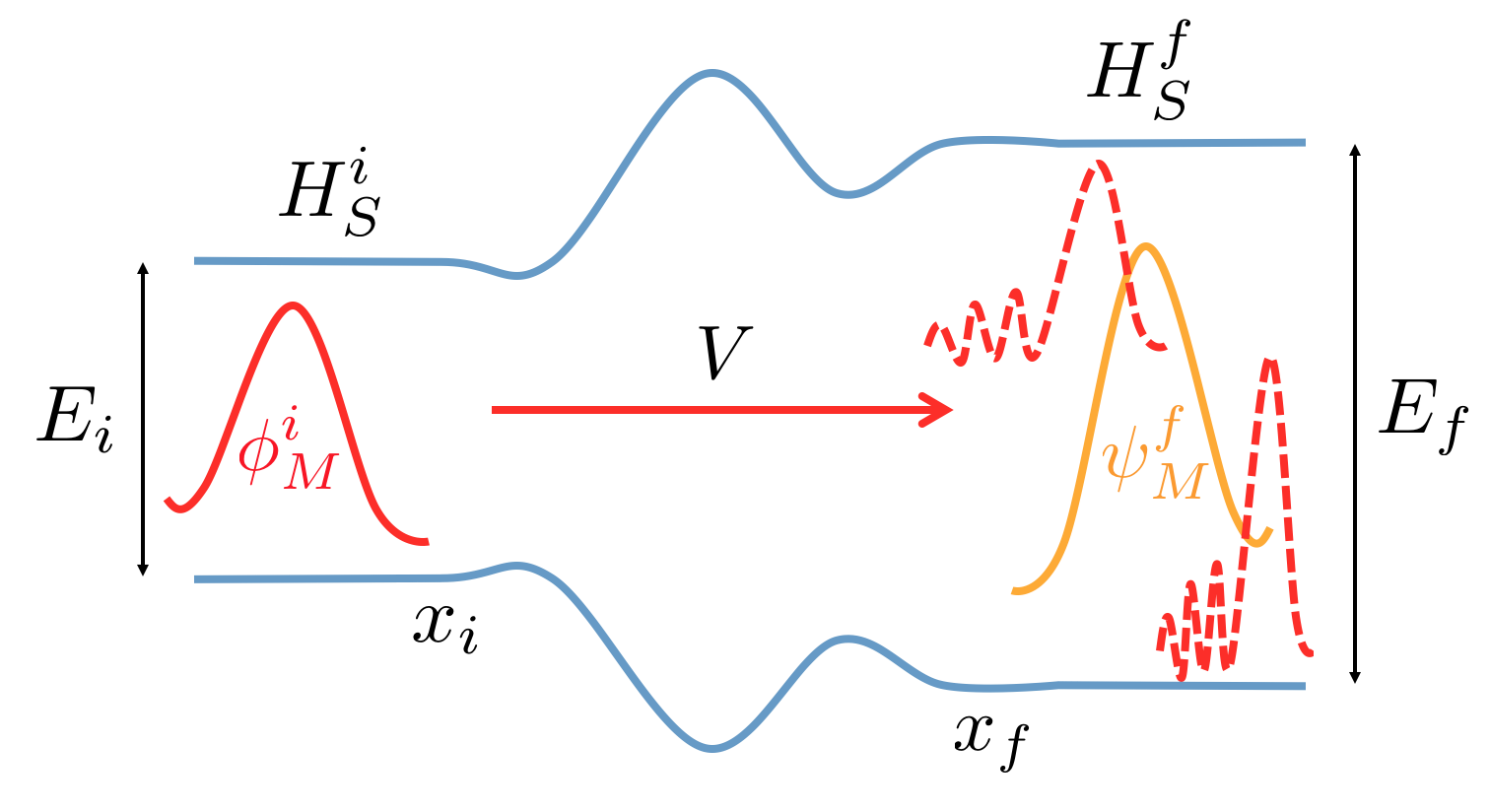}} 
\hspace{10mm}
\subfloat[Reverse Protocol]{\includegraphics[width = 2.7in]{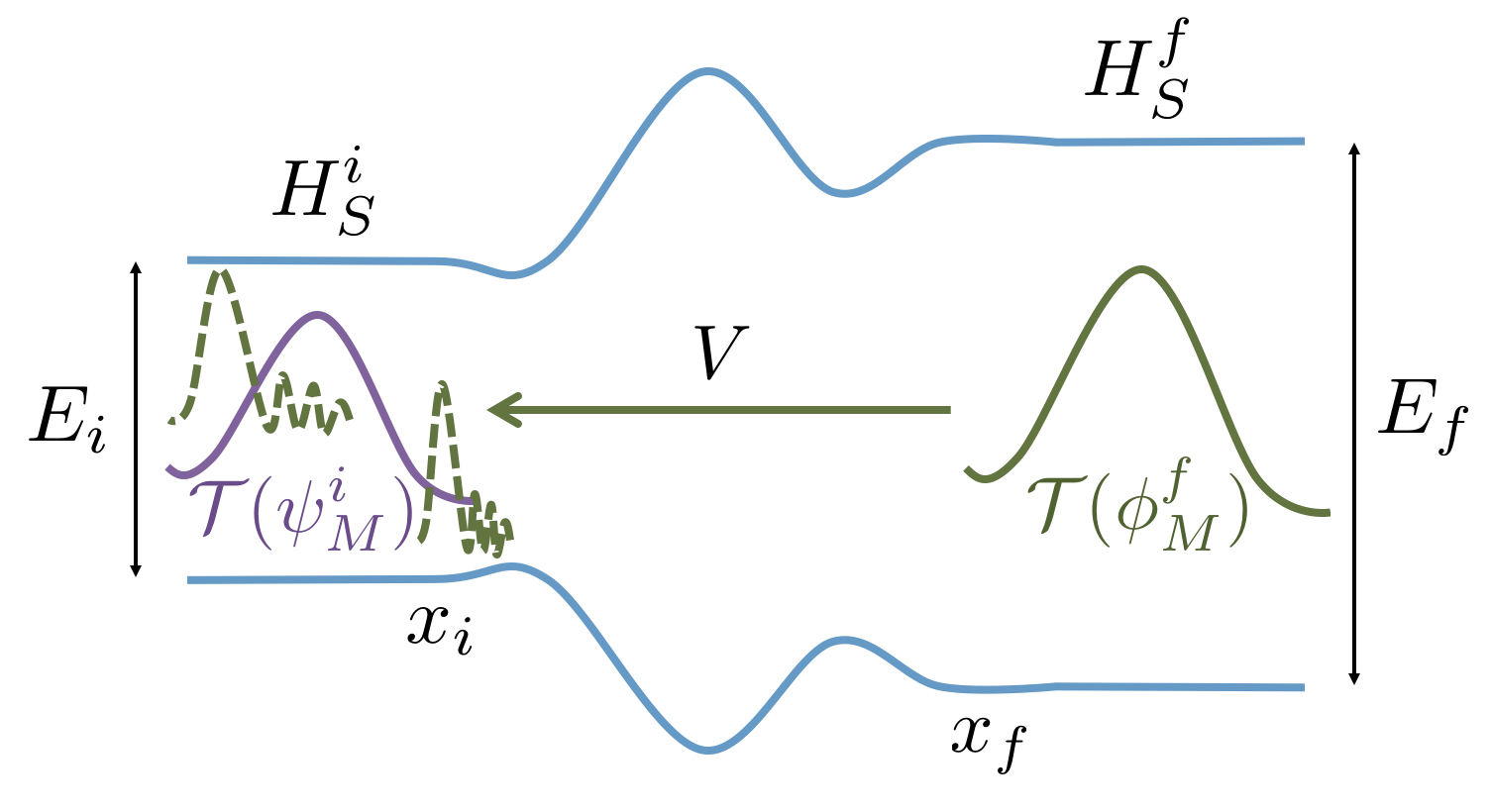}} 
\caption{\label{Fig: spin example sketch}This is a sketch of the forwards and reverse protocols respectively for the two level system and motional machine example. The blue lines represent the ground and excited state of the two-level system as a function of position. The solid lines represent the wave packets of the machine that are prepared at the start of the protocols (red/green) and the measurements (yellow/purple) that are performed at the end of the protocols. The dashed lines represent the evolved thermal machine states.}
\end{center}
\end{figure}

For concreteness, we can picture the bipartite setup using the example of a two level system and a motional thermal machine that interact via,
\be\label{eq: level splitting interaction}
H_{MS}^{\mbox{\tiny{int}}} =  E(x_M) \otimes \sigma^z_S \ .
\ee
$E(x_M)$ is an energetic level-shift that depends on the position, $x_M$, of the machine. $\sigma^z_S = \ket{e_S}\bra{e_S} - \ket{g_S}\bra{g_S}$ with $\ket{e_S}$ and $\ket{g_S}$ the excited and ground states of the two level system respectively. This setup could be physically implemented by a spin in a position dependent magnetic field. As sketched in Fig.~\ref{Fig: spin example sketch}, by choosing the function $E(x)$ to be constant for $x\le x_i$ and for $x\ge x_f$, two distinct effective Hamiltonians, $H_S^i$ and $H_S^f$, can be realised for the two level system. Similarly to the classical Crooks equality, the coherent Crooks equality is independent of precisely how the Hamiltonian changes from $H_S^i$ to $H_S^f$ and as such the choice of the form of $E(x)$ in the region $x_i < x < x_f$ is arbitrary. The position of the machine controls the Hamiltonian of the two level system and its energy provides the work store. 

\medskip 
The aim now is to use this framework to probe the energy fluctuations that result from driving a system with a coherent energy supply away from equilibrium. This is done by broadly replicating the standard classical Crooks protocol but now allowing the thermal machine to exist in a superposition of energy states. The new protocol, as sketched in Fig~\ref{Fig: spin example sketch}, consists of the following three stages:
\begin{description}[align=left]
\item[Preparation] The machine is prepared in some state $\ket{\phi_M^i}$ such that the initial effective Hamiltonian of the system is $H_S^i$ (i.e. for our example localised such that $\bra{\phi_M^i} x_M \ket{\phi_M^i} < x_i$). This is the only constraint on $\ket{\phi_M^i}$ which in general can be in a superposition of energy states. As in the classical case, the system is prepared in a thermal state, at temperature $T$, with respect to its initial Hamiltonian, $\gamma(H_S^{i}) \propto \exp\left(-\frac{H_S^{i}}{k_BT}\right)$.
\item[Evolution] The system and machine then evolve under an energy conserving unitary $V_{MS}$ that induces a change in the effective system Hamiltonian from $H_S^i$ to $H_S^f$. This drives the system from equilibrium with the energy required to do so supplied by the machine.
\item[Measurement] To quantify the energy changes of the driven system, a binary projective measurement is performed on the machine, \red{$\left\{\ket{\psi_M^f}\bra{\psi_M^f}, \I_M - \ket{\psi_M^f}\bra{\psi_M^f}\right\}$}. The only restriction on the state $\ket{\psi_M^f}$ is that it must correspond to the final system Hamiltonian $H_S^f$ (i.e. for our example be localised such that $\bra{\psi_M^f} x_M \ket{\psi_M^f} > x_f$). In general $\ket{\psi_M^f}$ is a superposition of energy states and so this measurement probes the coherent properties of the evolved system and machine.
\end{description}
The transition probability for the forwards protocol,
\red{\begin{equation}\label{Eq: ForwardsProb}
\mathcal{P}_{H_S^i \rightarrow H_S^f}\big(\ket{\phi_M^i} \rightarrow \ket{\psi_M^f}\big) := \Tr \left[\big(\ket{\psi_M^f}\bra{\psi_M^f} \otimes \I_S \big)V_{MS}\big(\ket{\phi_M^i}\bra{\phi_M^i} \otimes \gamma(H_S^{i})\big) V_{MS}^\dagger \right] \ , 
\end{equation}}
is the probability for the machine to collapse onto $\ket{\psi_M^f}$ having been prepared in $\ket{\phi_M^i}$. 

Similarly to the classical Crooks equality, the coherent Crooks equality purports to quantify the irreversibility of non-equilibrium processes. As such, the transition probability for the forwards process, Eq.~\eqref{Eq: ForwardsProb}, is compared to one for its time-reversed variant. The time reversal operation, $\T$, is introduced to characterise this reverse process. Specifically, $\T$ is defined such that if we now assume that the setup and its evolution are time reversal invariant, i.e. $\T(H_{MS}) = H_{MS}$ and $\T(V_{MS}) = V_{MS}$, we have that when the machine is prepared in some state\footnote{To avoid a proliferation of notation we use the symbol $\T$ to denote both a mapping on the level of Hilbert spaces and a map on the space of operators on the Hilbert space, i.e. here we abbreviate the time reversal of a pure state $\T (\ket{\psi} \bra{\psi} )$ to  $\T \ket{\psi}$.} $\T \ket{\phi_M^f}$ its evolution under $V$ drives a change in Hamiltonian from $H_S^f$ back to $H_S^i$. (We formally define $\T$ in Section III.) The transition probability for the reverse process,
\red{\begin{equation}\label{Eq: TimeReversedProb}
\mathcal{P}_{H_S^f \rightarrow H_S^i}\big(\T \ket{\phi_M^f} \rightarrow \T \ket{\psi_M^i}\big) := \Tr\left[\big(\T(\ket{\psi_M^i}\bra{\psi_M^i}) \otimes \I_S \big) V_{MS} \big(\T(\ket{\phi_M^f}\bra{\phi_M^f}) \otimes \gamma(H_{S}^f)\big)V_{MS}^\dagger \right] \ ,
\end{equation}}
is otherwise entirely analogous to the forwards case and quantifies the probability to find the machine in $\T\ket{\psi_M^i}$ having prepared it in $\T\ket{\phi_M^f}$. 

\medskip 
In the coherent Crooks equality, the pair of machine states quantified in the reverse process, $\T\ket{\psi_M^i}$ and $\T\ket{\phi_M^f}$, are constrained by the choice in states for the forwards process, $\ket{\psi_M^f}$ and $\ket{\phi_M^i}$.
The equality is derived by considering the ratio of the forwards, Eq~\eqref{Eq: ForwardsProb}, and reverse, Eq.~\eqref{Eq: TimeReversedProb}, transition probabilities and exploiting the fact that $H_{MS}$ and $V_{MS}$ commute and are time reversal invariant. It is found that the comparison of these probabilities is only possible for pairs of states related by a temperature dependent operation, 
\begin{equation}\label{eq:tempdepop}
\begin{aligned}
&\ket{\phi_M^i} \propto \exp\left(-\frac{H_M}{2k_B T}\right)\ket{\psi_M^i} \ \ \ \  \mbox{and} \\
&\ket{\phi_M^f} \propto \exp\left(-\frac{H_M}{2k_B T}\right)\ket{\psi_M^f} \ .
\end{aligned}
\end{equation}
For energy eigenstates this operation is trivial; we have that $\ket{\phi_M^{i,f}} = \ket{\psi_M^{i,f}}$ and the pairs of states in the forwards and reverse processes are simply the time reverse of one another. In general, the operation is non-trivial but can be seen as emerging from a seemingly natural map, called the Gibbs map, that we discuss in Section IV.

Having parameterised the relevant machine states in this way, the ratio of the transition probabilities is calculable and the coherent Crooks equality is found,
\begin{equation}\label{Eq: Quantum Crooks}
\begin{aligned}
&\frac{\mathcal{P}_{H_S^i \rightarrow H_S^f}\left(\ket{\phi_M^i} \rightarrow \ket{\psi_M^f}\right)}{\mathcal{P}_{H_S^f \rightarrow H_S^i}\left(\T\ket{{\phi}_M^f} \rightarrow  \T\ket{\psi_M^i}\right)} = \exp \left(\frac{1}{k_B T}\left(\Delta \tilde{E} - \Delta F\right) \right) \ .
\end{aligned}
\end{equation}
(This equality is Eq. 28 of~\cite{aberg} but written for a bipartite setup and assuming that the machine is prepared in a pure state.)
The dependence on the change in equilibrium free energy $\Delta F$ carries over from the classical Crooks equality~\cite{Crooks}, where as usual $\Delta F := F( H_S^f, \, T) - F( H_S^i, \, T)$ with the free energy for any Hamiltonian $H$ and temperature $T$ given by
\begin{equation}\label{eq:free energy}
\begin{aligned}
F( H, \, T) := -k_B T \ln \left( \Tr \left[\exp \left(-\frac{H}{k_B T}\right)\right] \right) \ .
\end{aligned}
\end{equation}
However, the classical work term is replaced with a quantum generalisation of the energy supplied to, or absorbed from, the system, $\Delta \tilde{E}:=  \tilde{E}_{\ket{\psi_M^i}\bra{\psi_M^i}}( H_M, \, T) - \tilde{E}_{\ket{\psi_M^f}\bra{\psi_M^f}}(H_M, \, T)$. The function $ \tilde{E}_{\rho}$ is a state dependent mathematical generalisation of the equilibrium free energy defined as
\begin{equation}\label{eq:flow}
\begin{aligned}
&\tilde{E}_\rho( H, \, T )  := -k_B T \ln \left(\Tr \left[ \exp\left(- \frac{ H}{k_BT}\right) \rho\right]\right)\ .
\end{aligned}
\end{equation}
As required to regain the classical limit, when the thermal machine is prepared in an energy eigenstate $\Delta \tilde{E}$ corresponds to the energy exchanged between the system and the thermal machine. More generally, as we discuss in Section IV, $\tilde{E}$ appears to be a natural, temperature dependent, quantum energy measure.


The deviations from the classical Crooks equality are encapsulated in these two new concepts: the temperature dependent operation that parameterises the relevant quantum states and the substitution of the classical work term for $\Delta \tilde{E}$. For energy eigenstates $\{ \ket{E_M^k} \}$ and corresponding energies $E_k$, the coherent Crooks equality reduces to 
\begin{equation}\label{Eq: classical limit quantum crooks equality}
\begin{aligned}
&\frac{\mathcal{P}_{H_S^i \rightarrow H_S^f}\left(\ket{E_M^i} \rightarrow \ket{E_M^f}\right)}{\mathcal{P}_{H_S^f \rightarrow H_S^i}\left(\ket{E_M^f} \rightarrow \ket{E_M^i}\right)} =  \exp \left(\frac{1}{k_B T}\left(\left(E_i-E_f\right) - \Delta F\right) \right) \ .
\end{aligned}
\end{equation}
If we then additionally assume that the evolution of the setup does not depend on the initial energy of the machine, the usual classical Crooks equality is regained by identifying the change in energy of the machine with the work done on the system. It is primarily in virtue of this that the coherent Crooks equality can be seen as a genuine quantum generalisation. However, in general, the temperature dependent parameterisation of the relevant quantum states and the generalised energy flow term are essential to capture the impact of coherence.

\section{III. \ Conceptual Ingredients}

\paragraph*{Control system.}
The example of the two level system that experiences a position dependent splitting is intended to provide an intuition as to the role of the control. More generally, the system Hamiltonian is changed by moving the state of the control between two regions of the total Hilbert space corresponding to different effective system Hamiltonians.  This is sketched in Fig.~\ref{Fig: sketch of equality plot}b. Consider a system coupled to the control by a Hamiltonian of the form
\begin{equation}\label{eq: hilbert space}
H_{CS} =  \Pi_{C}^i \otimes H_{S}^i +  \Pi_{C}^f \otimes H_{S}^f + H_{CS}^\perp 
\end{equation}
where $\Pi_{C}^i$ and $\Pi_{C}^f$ are projectors onto different subregions, $R_i$ and $R_f$ respectively, of the control. $H_{CS}^\perp$ has support only outside those two subspaces, i.e. \red{$(\Pi_C^i \otimes X_S)H_{CS}^\perp=(\Pi_C^f \otimes X_S)H_{CS}^\perp=0$} for any system operator $X_S$. When the control is prepared in a state with support in region $R_i$ only, the initial effective system Hamiltonian is $H_{S}^i$. \red{The evolution of the control system is induced by a unitary operation that is chosen to switch the control system from a state with support in region $R_i$ to one in $R_f$}. This changes the effective system Hamiltonian to $H_S^f$. 
\medskip

\paragraph*{Time reversal.} 
As the classical Crooks equality relates a forwards process to its time reversed variant, any quantum generalisation of it requires a means of defining this reversed process. For this reason, the quantum time reversal operation $\T$ is introduced.

The time reversal operation can be enacted by complex conjugation or the transpose operation. While complex conjugation is the `textbook' \cite{timereversalop} quantum time reversal operation, its anti-linearity can make it mathematically arduous and so in \cite{aberg} the transpose is used. The two are equivalent on Hermitian operators and as such we have that for any state or observable, $\sigma$, that $\T(\sigma) :=\sigma^T \equiv \sigma^*$ where $^*$ and $^T$ are the complex conjugation and transpose operations respectively. The basis that the transpose is taken with respect to is dictated by the physical implementation considered. Specifically, it is chosen such that the time reversed control states drive the change in Hamiltonian from $H_S^f$ back to $H_S^i$ and such that the global Hamiltonian and evolution operator are time reversal invariant.

A sense of how $\T$ operates can be gained by looking at its effect on the evolution of a state. Suppose an initial state $\rho_i$ evolves under a time reversal invariant unitary, $U = \T(U) = U^T$, to some state $\rho_f$, i.e. $\rho_f = U \rho_i U^\dagger$. Then the time reversed version of the final state $\T(\rho_f)$ evolves back under $U$ to the time reversed version of the initial state $\T(\rho_i)$, i.e. $\T(\rho_i) = U \T(\rho_f) U^\dagger$. This follows from rearranging $\rho_f^T = {U^\dagger}^T \, \rho_i^T \, U^T = U^\dagger \, \rho_i^T \, U$. 

In terms of our bipartite example, it is natural to take $\T()$ to be the transpose operation in the $x_M$ and $\sigma_S^z$ bases. This ensures the thermal states of the two level system are invariant, but that the momentum of the motional states are reversed, under $\T$. Suppose, for example, that in the forwards process the motional state is prepared and measured in optical coherent states~\cite{coherentstatesreference} $\ket{\alpha_i}\bra{\alpha_i}_M$ and $\ket{\alpha_f}\bra{\alpha_f}_M$, centered in the regions $x < x_i$ and $x > x_f$ respectively and with momentum in the positive $x$ direction. Then the states for the reverse process, leaving aside the additional impact of the temperature dependent operator in Eq~\eqref{eq:tempdepop}, are $\ket{\alpha_i^*}\bra{\alpha_i^*}_M$ and $\ket{\alpha_f^*}\bra{\alpha_f^*}_M$. As the position and momentum of a coherent state $\ket{\alpha}$ are proportional to the real and imaginary parts of $\alpha$ respectively, this reverses the momentum of the \red{machine} while leaving its position unchanged. As such, preparing the \red{machine} in $\ket{\alpha_f^*}\bra{\alpha_f^*}_M$ drives the required reverse change in effective system Hamiltonian.

\medskip

\paragraph*{Derivation.}
The cleanest derivation of the coherent Crooks equality makes use of two properties: `global invariance' and `factorisability'. 

Global invariance is a property of a pair of states, $\rho_i$ and $\rho_f$, of a
single system with a time reversal invariant Hamiltonian $H$ and a strictly energy conserving and time reversal invariant \red{unitary} evolution $V$. By exploiting the fact that $H$ and $V$ are time reversal invariant and commute it is found that the quantity
\begin{equation}
\Tr\left[\rho_f V \exp\left(-\frac{H}{2k_B T}\right) \rho_i \exp\left(-\frac{H}{2k_B T}\right) V^\dagger\right] 
\end{equation}
 is invariant under the transformation $\rho_i \rightarrow \T(\rho_f)$ and $\rho_f \rightarrow \T(\rho_i)$. This property is the starting point to derive a large family of quantum fluctuation theorems and plays an analogous role to detailed balance for classical fluctuation theorems. 
 
Factorisability characterises the extent to which multiple interacting systems can be considered independent subsystems and enables a Crooks-like equality to be derived from global invariance. This condition holds when the system, thermal bath, control and work store are effectively non-interacting at the start and end of the forwards and reverse protocols. Specifically, the bipartite setup is factorisable if the exponential of $H_{MS}$ factorises into a $H_M$ term and a $H_S^{i}$ or $H_S^{f}$ term when acting on the machine states $\ket{\psi_M^{i}}$ and $\ket{\psi_M^{f}}$ respectively, i.e. if
 \begin{equation}\label{Eq: factorisability}
 \begin{aligned}
 &\exp\left(-\frac{H_{MS}}{2k_B T}\right) ( \ket{\psi_M^k}\bra{\psi_M^k} \otimes \I_S ) \exp\left(-\frac{H_{MS}}{2k_B T}\right) \\ &= \exp\left(-\frac{H_{M}}{2k_B T}\right) \ket{\psi_M^k}\bra{\psi_M^k} \exp\left(-\frac{H_{M}}{2k_B T}\right)   \otimes \exp \left( - \frac{H_S^k}{k_B T}\right)
 \end{aligned}
 \end{equation}
 for $k = i$ and $k = f$.

For a complete derivation see Appendix~I of~\cite{aberg} or~\cite{mypaper,erick}. 

\medskip 

\paragraph*{Externally controlled and autonomous variants.}

There are two variants of the coherent Crooks equality corresponding to different ways in which the evolution \red{is} induced. The equalities take the same general form; however, they are derived from different assumptions about the setup Hamiltonian.

In the externally controlled variant, the evolution is induced by the \textit{application} of some energy conserving unitary operation $V$. It is more natural to picture this variant when the control and work store are independent systems. The coherent Crooks equality holds exactly for this external controlled variant as long as the system and work store Hamiltonians are non-interacting\footnote{In the bipartite case the thermal machine and system Hamiltonians must interact for the machine to act as the control. For the coherent Crooks equality to hold exactly, the machine Hamiltonian is instead required to not induce evolution between the initial and final subregions, i.e. $ (\I_M - \Pi_M^i ) H_M \Pi_M^i = 0$ and $ (\I_M - \Pi_M^f ) H_M \Pi_M^f = 0$.}, i.e.
\begin{equation}
H_{WCS} =  \I_W \otimes H_{CS}  + H_{W} \otimes \I_{CS} \ ,
\end{equation}
where $H_{CS}$ is defined in Eq.~\eqref{eq: hilbert space}. However, there is an apparent tension here. The time evolution of the setup will be determined by some Hamiltonian $H_{WCS}^{\mbox{\tiny evol}}$ where $V_{WCS} = \exp(- i H_{WCS}^{\mbox{\tiny evol}} t)$. This Hamiltonian must contain interaction terms to enable the exchange of energy between the system and work store. As such, $H_{WCS}^{\mbox{\tiny evol}} \neq H_{WCS}$ and we have two distinct Hamiltonians for the setup. This tension can be reconciled by thinking of $H_{WCS}$ as parameterising the states at the start and end of the protocols and $H_{WCS}^{\mbox{\tiny evol}}$ as the `true' Hamiltonian that completely describes the energy of the setup.

The second variant avoids the tension entirely; however, at the cost of not being exact. In this case the setup Hamiltonian includes a term to drive the evolution of the control and to transfer energy between the system and work store. The setup then evolves \textit{autonomously} under this Hamiltonian. This case is inexact because in order for the autonomous evolution of the setup to induce the required change in system Hamiltonian, the system must be interacting with the work store and control at all times. As a result the factorisability condition does not hold exactly. The approximate nature of this autonomous variant is formally quantified by an error bound in~\cite{aberg}. Numerical simulations~\cite{aberg,mypaper} indicate that for a wide parameter range, the errors fall below plausible experimental error margins. Hence, in such regimes these equality can nonetheless essentially be treated as exact.

The two level system with a motional machine example is most naturally understood in the autonomous variant. In this example, the machine Hamiltonian would simply contain kinetic, and possibly potential, energy terms. For example, $H_M$ in Eq.~\eqref{eq: total hamiltonian} could be a harmonic oscillator Hamiltonian centered halfway between $x_i$ and $x_f$. When the system and thermal machine evolve under $V_{MS} = \exp \left(- i H_{MS} t \right)$ this then drives the required translation process. The coherent Crooks equality holds to a high degree of accuracy when the machine is prepared far from interaction region, i.e. outside of the region $x_i < x <x_f$ in Fig.~\ref{Fig: spin example sketch}. This is intuitive because when prepared far from the interaction region, the interaction between the machine and the two level system is effectively negligible and consequently they can be seen as approximately independent.

The control in the autonomous variant plays a role similar to that of the quantum clocks that have been studied elsewhere in quantum thermodynamics~\cite{clocksthermo} and quantum information theory~\cite{clocks1,clocks2}. As with quantum clocks, this variant is partially motivated by the desire to avoid the implicit dependence on an additional, potentially classical, system (the experimentalist and their apparatus) that is required to apply a unitary operation.

\section{IV. \ Conceptual Significance} \label{sec:conceptual}
\paragraph*{Gibbs map and generalised energy flow.}
We saw in Section II that two new concepts emerged from applying the standard Crooks equality approach of comparing a forwards and reverse process in the presence of a coherent energy supply: firstly, the operation that parameterises the thermal machine states, Eq.~\eqref{eq:tempdepop}, and secondly, the function $\Delta \tilde{E}$ that replaces the classical work term in the coherent Crooks equality, Eq~\eqref{Eq: Quantum Crooks}. Their precise forms, in particular their temperature dependence, are forced by the derivation of the coherent Crooks equality. While the full significance of these new concepts is very much an open research question; a study of their basic properties suggests that they are not only convenient mathematical definitions but also physically natural.

\medskip

The temperature dependent operation that parameterises the machine states in Eq.\eqref{eq:tempdepop}, emerges from the Gibbs map, 
$G_\rho$, which given a system with Hamiltonian $H$ at temperature $T$, is defined as
\begin{equation}\label{eq:gibbsmap}
\begin{aligned}
&G_\rho(H, \, T) := \frac{\exp\left(-\frac{H}{2 k_B T}\right) \rho \exp \left(-\frac{H}{2 k_B T}\right)}{\tilde{Z}_\rho(H, \, T)} \ , \\  
&\tilde{Z}_\rho(H, \, T) := \Tr\left[\exp\left(-\frac{H}{ k_B T}\right) \rho \right] \ .
\end{aligned}
\end{equation}

This map arises naturally as a quantum-mechanical version of the Crooks reversal of a Markov process, and is intimately linked with the Petz recovery map for general quantum states~\cite{petz,hyukjoon}. An intuition as to the action of the map can be obtained through a couple of examples. 
When the energy is exactly known, the state is completely constrained, and the map has no effect: an energy eigenstate is left unchanged. 
However, when the energy of the state is completely uncertain, as in a maximally mixed state, $\propto \I$,  or in an equal superposition, $\propto \sum_k^N \ket{E_k}$, a thermal rescaling is applied such that the maximally mixed state is mapped to the thermal state and the equal superposition to the coherent thermal state, a pure state with the same energy populations as the thermal state i.e. $\propto \sum_k \exp\left(-E_k / 2 k_B T \right)\ket{E_k}$.
Note, the map is non-dephasing and affects the energy populations in the same way irrespective of the coherent properties and phase of the state.
In this way, the map makes a state crudely `as thermal as possible' subject to the constraints imposed by the input state and the temperature of the bath. However, this loose claim is not intended to be taken literally but rather as a signpost towards the map's deeper physical significance.

\medskip

The energies of the machine states are parameterised by $\tilde{E}$ as defined in Eq.~\eqref{eq:flow}. This function is a state dependent mathematical generalisation of the equilibrium free energy, Eq~\eqref{eq:free energy}, in which the standard partition function, $Z(H, T) := \Tr\left[\exp\left(-\frac{H}{k_B T}\right)\right] $, is replaced by the Gibbs map normalisation term $\tilde{Z}_\rho(H, T)$.

A study of the properties of $\tilde{E}$ hints at its naturalness as a statistical, temperature dependent, energy scale for quantum states. For an energy eigenstate, $\tilde{E}$ is simply the associated eigenstate energy. This is intuitive because when the energy of a state is well-defined there is no need to statistically estimate it. Moreover, it is required to regain the classical limit.
More generally, $\tilde{E}$ is upper bounded by the average energy $\langle H \rangle$, and tends towards it in the high temperature limit. 
In the general finite temperature quantum case, where $\tilde{E}$ does not coincide with the average energy, the function nonetheless obeys several physically desirable properties for an statistical energy scale. Firstly, $\tilde{E}$ scales with a constant offset or multiplicative factor to $H$ as one would expect an energy measure to scale; i.e. $\tilde{E}_\rho(H + \delta,\, T) = \tilde{E}_\rho(H,\, T) + \delta$ and $ \tilde{E}_\rho(\lambda H, \, T) = \lambda \tilde{E}_\rho(H, \, \lambda T)$. Secondly, $\tilde{E}$ depends on the energy populations of a state only. This means that (i) it is independent of the phase of the state and as such in the absence of interactions remains constant in time and (ii) it takes the same value for a pure state and its completely dephased variant. Again, these properties imply that $\tilde{E}$ has deeper physical significance that is worth further investigation. 

\textit{Note added: since publishing this chapter $\tilde{E}_\rho$ has been identified as the cumulant generating function~\cite{cumulant} for the measurement statistics of energy in the quantum state $\rho$ and an account has been developed in which $\tilde{E}_\rho$ is interpreted as an effective potential with both energetic and coherent contributions~\cite{erick} .}

\medskip 

\paragraph*{Thermal operations link.}
Resource theories take properties that are in some sense useful, but usually scarce, and attempt to precisely characterise them. They do this by specifying a restricted set of operations, known as `free operations', and defining the states that cannot be generated using solely free operations as `resources'. Of particular relevance to quantum thermodynamics are the resource theories of entanglement~\cite{entanglementresource}, coherence~\cite{coherenceresource}, noise~\cite{noiseresource} and the thermal operations framework~\cite{chap25,resourcetheory,resourcetheory1,resourcetheory3}.

The quantum channel induced on the machine in the coherent Crooks equality is a thermal operation and as such previous results within the thermal operations framework~\cite{resourcetheory1,resourcetheory3,resourcetheory,completestateinterconversion, 2ndlaws,constraintsbeyondfreeenergy, workextractionaberg,catalyticcoherence, catalyticworkextraction} are applicable. Thermal operations, consist of the following three operations: firstly, performing a global unitary, $U$, that strictly conserves total energy, $H$, in the sense that $[H, U] = 0$; secondly, adding a system at thermal equilibrium; and thirdly, disregarding (tracing out) any system. The quantum channel on the machine is precisely of this form. For the forwards process we have,
\begin{equation}
\E (\rho_M) = \Tr_{S} \left[ V_{MS} (\rho_M \otimes \gamma(H_S^{i})) V_{MS}^\dagger \right] \ , 
\end{equation}
and the reverse process channel is identical except the system is prepared in a thermal state with respect to its final Hamiltonian. 
The resource states of the thermal operations framework are states with coherence with respect to the energy eigenbasis or non-thermally distributed populations. Correspondingly, the machine states are resources in virtue of being prepared in a pure state. For further discussion on incorporating fluctuating work into the thermal operations framework see~\cite{alvaro}.

\medskip

\paragraph*{Off-diagonal equality.}
\begin{figure}
\begin{center}
\subfloat[]{\includegraphics[width = 2.7in]{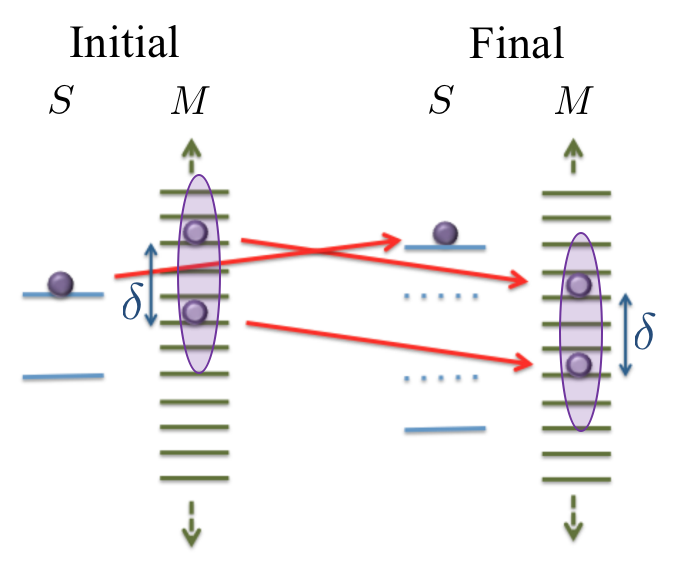}} 
\hspace{12mm}
\subfloat[]{\includegraphics[width = 2.7in]{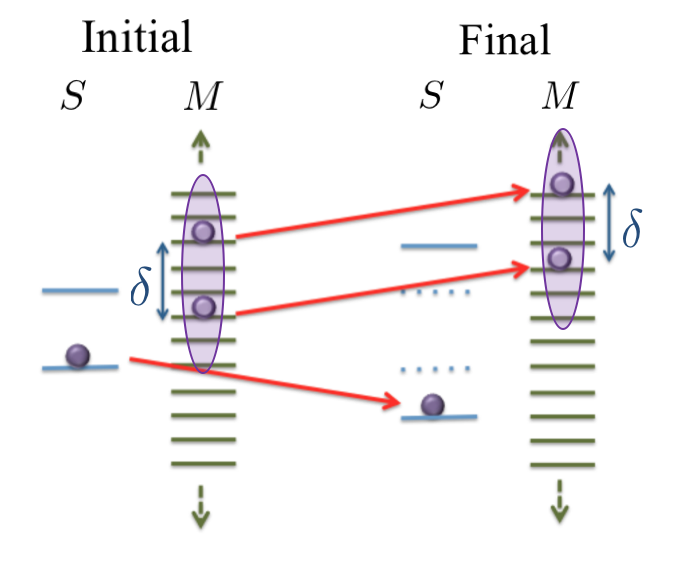}} 
\caption{\label{Fig: offdiagonal} The process of increasing the splitting of a two level system in the presence of a harmonic oscillator machine. The left and right figures cover the cases when the qubit system starts in the excited (a) and ground (b) states respectively. The see-saw like mechanism, highlighted by the red arrows, ensures energy conservation and maps changes in the system energy onto that of the machine. A superposition of machine states is raised or lowered `as one' and as such the coherences evolve in the same manner as the populations.}
\end{center}
\end{figure}

The form of the coherent Crooks equality can be better understood via a closely related equality also derived in~\cite{aberg}: the off-diagonal Crooks equality. This equality states that the evolution of the energy coherences of the thermal machine (the off-diagonal elements of thermal machine density operator with respect to the energy eigenbasis), evolve in the same way as the populations (the diagonal elements). The evolution of the off-diagonal elements of the machine in the energy basis are quantified by the transition amplitudes $q_{+}^\delta(f|i) := \bra{E_M^f} \Tr_S[ V (\ket{E_M^i}\bra{E_M^{i+\delta}}\otimes \gamma(H_{S}^f)) V^\dagger ] \ket{E_M^{f+\delta}}$ and $q_{-}^\delta(i|f) := \bra{E_M^{i+\delta}} \Tr_S[ V (\ket{E_M^{f+\delta}}\bra{E_M^{f}}\otimes \gamma(H_{S}^f)) V^\dagger ] \ket{E_M^{i}}$. Similarly to the diagonal elements in Eq.~\eqref{Eq: classical limit quantum crooks equality}, they are constrained to obey
\begin{equation}\label{Eq:off diag crooks eq} 
\frac{q_+^\delta(f|i)}{q_-^\delta(i|f)} = \exp \left(\frac{1}{k_B T}\left((E_i-E_f) - \Delta F\right) \right) \ .
\end{equation}

The off-diagonal Crooks equality can be seen as emerging from the non-trivial constraints imposed by thermal operations. That the coherences are constrained to evolve in the same way as the populations is a result of strict energy conservation and the fact the system is initially thermally distributed. Pictorially, as sketched in Fig.~\ref{Fig: offdiagonal}, energy conservation enforces a see-saw operation that maps changes in the system energy onto that of the machine. The strict conservation of energy then further ensures that a superposition of machine states is raised or lowered `as one' and as such the coherences evolve in the same manner as the populations. Note, these constraints imposed by energy conservation can also be seen as resulting from time translation symmetry.

\medskip 

\section{IV. \ Conclusions}

We have given a brief summary of a framework introduced in~\cite{aberg,alvaro} which endeavours to extend work fluctuations to the quantum regime using ideas from the information theoretic approach to quantum thermodynamics. This approach is distinguished by its use of time independent Hamiltonians and its decision to explicitly model not only a driven system and its thermal bath but also the control system and work store that enable the system to be driven. Specifically, we have presented one of the key results of this framework: a Crooks-like equality for a system with a coherent energy supply, Eq.~\eqref{Eq: Quantum Crooks}. 

The equalities that we have presented in this chapter are not the most general. There are variants of Eq.~\eqref{Eq: Quantum Crooks} in which (to name just a few): (i) the control and work store are separate systems; (ii) the work store is prepared in a mixed state and correspondingly a non-projective POVM is performed at the end of the protocol. (iii) the system is prepared in a non thermal state. (iv) entanglement between the subsystems is incorporated. For these more general cases and others see~\cite{aberg}. 

The thermal bath can play a more active role than it does in the version we have discussed. The simplest way of doing so is to reinterpret the system $S$ that we have been discussing as an enlarged system that incorporates the thermal bath. This amounts to effectively dealing with some large system that starts at equilibrium and considering changing the Hamiltonian of a small part of it. Alternatively, a quantum fluctuation relation for Markovian master equations is included in~\cite{aberg} to provide a link to the open quantum systems approach to quantum thermodynamics.

Moreover, in~\cite{aberg} the quantum crooks equality is initially formulated in terms of the forwards and reverse quantum channels induced on the work store. This is more in the spirit of a fully quantum, information theoretic, Crooks equality. For the purpose of a pedagogical introduction, we chose to present its reformulation in terms of transition probabilities in order to maintain a closer resemblance to the classical Crooks equality. 

An advantage of this transition probability formalism is that it is clearer how one might go about experimentally testing the coherent Crooks equality. This is especially true of its autonomous variant. The example of the two level system that experiences a position dependent splitting is not just a pedagogical example but the basis of an experimental proposal utilising trapped ions and the AC Stark shift that is presented in~\cite{mypaper}. As such while the equality might appear rather abstract, it is nonetheless physically implementable.

The coherent Crooks equality is conceptually new and there remain numerous open research questions. As already discussed, the physical significance of the Gibbs map and the generalised energy flow have yet to be pinned down. In~\cite{aberg,alvaro,hyukjoon,erick}, links are touched on between the coherent Crooks equality and other key concepts in quantum information theory including the Petz recovery map~\cite{petz}, quantum reference frames~\cite{referenceframe} and quantum clocks~\cite{clocks1, clocks2, clocksthermo}; these warrant further investigation. There is also the question of whether the equality could be converted from its current form in terms of transition probabilities of the energy supply, to an equality that is stated in terms of probability distributions of the energy changes of the system. More fundamentally, while it is clear that quantum coherence generates non-trivial quantum corrections to these fluctuation theorems, questions remain as to the implications for our understanding of irreversibility in quantum mechanics and as to whether they could provide any practical quantum advantage. 
\bigskip

\acknowledgements

ACKNOWLEDGEMENTS
The author thanks Johan \AA berg, \'{A}lvaro Alhambra, Janet Anders, Florian Mintert, Felix Binder, Erick Hinds Mingo, Tom Hebdige and Jake Lishman for commenting on drafts and David Jennings for numerous indispensable discussions.

\bibliography{refs}

\end{document}